\documentclass[letterpaper,twocolumn,10pt]{article}
\usepackage{usenix,epsfig,endnotes}

\begin{document}

\date{}

\title{\Large \bf Blip: JIT and Footloose On The Edge}

\author{
{\rm Andy Edmonds}\\
Zürich University for Applied Sciences
\and
{\rm Chris Woods}\\
Mindflip Ltd.
\and
{\rm Ana Juan Ferrer}\\
Atos 
\and
{\rm Juan Francisco Ribera }\\
Zürich University for Applied Sciences
\and
{\rm Thomas Micheal Bohnert}\\
Zürich University for Applied Sciences
}

\maketitle

\section*{Abstract}
Edge environments offer a number of advantages for software developers including the ability to create services which can offer lower latency, better privacy, and reduced operational costs than traditional cloud hosted services. However large technical challenges exist, which prevent developers from utilising the Edge; complexities related to the heterogeneous nature of the Edge environment, issues with orchestration and application management and lastly, the inherent issues in creating decentralised distributed applications which operate at a large geographic scale.  In this conceptual and architectural paper we envision a solution, Blip, which offers an easy to use programming and operational environment which addresses the these issues. It aims to remove the technical barriers which will inhibit the wider adoption Edge application development. This paper validates the Blip concept by demonstrating how it will deliver on the advantages of the Edge for a familiar scenario.

\section{Introduction}
\label{intro}
Edge computing has opened up a range of new capabilities to developers. These are distributed along the length and breadth of the network from cloud data centres through to user devices, and a range of network and compute elements along the way. It also enables applications to structure and communicate in both a traditional data centre to client way (North and South), and should in a peer-to-peer way (East and West). This provides many advantages, including lower cost, higher privacy, data locality, and lower latency applications. This all sounds great, however a developer writing applications for this environment requires addressing many challenges. This includes:

\begin{itemize}
\item Heterogeneity of the fog environment, from embedded devices through to large servers
\item Orchestration and management of software across vast geographic areas
\item Service location and configuration; which communication can be North, South, East and West, and the same service can exist in any of these directions, knowing how to find and locate a service can be a challenge.
\end{itemize}

To be more succinct this paper asks: \textit{“Given the complexity of the Edge world, how do we create a programming environment where developers can easily create solutions which take advantage of having their code and applications hosted as close as possible to the user upon suitable resources?”}

By exploring this, we address the concerns outlined above. In this paper we present Blip, which allows a developer to create an application composed of workload element-based (a Blip concept) services, each of which can be scaled across the Edge, from data centre to user equipment. It addresses the complex issues of heterogeneity, and  orchestration. It presents a programming model based on a simple recursive architecture which allows developers to easily create applications, which can be decentralised and distributed across the large geographic areas which form the Edge.

This will allow developers to take advantage of the Edge, offering new applications and services. A common approach, software infrastructure, and programming model will allow multiple services from multiple vendors to be hosted in the Edge that offers new opportunities for telecommunication, CDN and cloud providers.

\section{Problems with the Cloud and Edge?}
\label{problem}
The Edge environment in inherently more heterogeneous than any other distributed environments for developers. There are two key reasons for this, firstly the hardware platform changes between various locations in the Edge, from ARM machines at the edge, through to x86 and vector acceleration in data centres. Secondly, the software infrastructure changes between providers. Developing efficient applications for cloud is hard alone. Considering an environment in which there are multiple vendors, like Edge creates further issues for developers to ensure that their solutions run at reasonable costs and efficiently. 

Creating applications which span the Edge will mean deploying code, and caching data across the network at the various available locations. This will require a cross-platform orchestration and life cycle management tool, which as yet doesn’t exist. 

Assuming the heterogeneity and the orchestration problems are resolved and a developer can write code that can be deployed and managed easily across the network, a final and important challenge remains. In order to leverage the benefits offered by the Edge environment (privacy, data locality, cost, and low latency) the developer will need to write applications that can execute in a truly distributed way, without a central focus point. This way, as many functions of the application as possible can be executed as close to the user as possible.

Many of the scenarios covered in the 5G Vertical Sectors paper \cite{c71} are valid targets for Blip and should be supported. Currently, we have considered the scenarios of Smart Home, Smart Industrial Plant (Industry 4.0) and Intelligent Transportation.

Digging deeper into the Smart Home scenario, the potential and challenges in the use of Blip can be understood. A common scenario is to control home lights with the software-enabled light bulbs (e.g. Philips Hue). These lights have wireless controllers communicating with a bridge. In the case that the user that wishes to actuate a lights and that user is connected to their local network, the response of the system is acceptable and near-real time. However when the system is used over current mobile networks (e.g. as envisioned in 5G white papers), the response is noticeably slow. In order to understand the slow response we need to understand the setup of the system and the path of the actuation request.

\textbf{System Setup}. The management of light bulbs is carried out by a bridge with a wired connection to the home router and a wireless connection to the light bulbs (e.g. Zigbee, or Wifi). The bridge can hold a HTTP websocket connection used to decrease latency [e.g. PHIL17] open to the light control application hosted on a regular cloud provider (CP). This websocket connection runs from the bridge, through the router out over the end-users Internet provider’s network, across the ISP’s peers and eventually routed to the light control application on the CP.

\textbf{Request Path}. To manage lights over the mobile network, an actuation request is issued. This first has to travel through the mobile device’s networking stack, be converted into a radio signal and then transmitted to the nearest mobile base station (C-RAN \cite{c73}). From here it enters into a system that typically adheres to the Evolved Packet Core architecture (EPC) \cite{c72}. Within this system there is a lot of encapsulation and decapsulation of tunneled packets (comprising the request), resulting (after policy and charging controls) in the exit of the packets into the mobile operator's Internet backbone. The request is then routed through a set of routers to arrive at the border of CP network. From here the request will go through a number of internal routers to arrive at the application instance responsible for serving the request. From here, the light control application issues an event to the bridge’s HTTP websocket. Once the event is received and processed by the light’s bridge, the command to turn off the light is issued from the bridge over the wireless medium, received by the light bulb which then turns off the LED within the bulb, resulting in the light being actuated.

\section{Architecture}
\label{principles}
In proposing Blip, we outline the set of founding principles, which can be used by Blip and its applications to respond to decentralised dynamic environments:

\textbf{Workload Element}: A compute component which is small that starts fast to support the Just-in-Time principle. Also the workload associated  must be small. We consider that both Unikernel-based VMs (also known as "library OSs") and Webassembly to be excellent candidates. Unikernels are typically within 3-10Mb in size and can start within 30ms \cite{c27}, and Webassembly is producing code which can execute at near to native speed \cite{c74}, while starting within almost as soon as the code has been downloaded. We expect that workload elements would “\textit{blip}” into existence and vanish once no longer needed. 

\textbf{Just-in-Time Services}. Services can be instantiated within milliseconds. The ability to start services only when requested, monitoring their usage and shutting them down when not longer needed. Offers low-latency invocation with zero over provisioning. JIT services provide two associated benefits: 1) cost reduction as services are only executed when needed. 2) if the service doesn't exist until need or if an attacker does not have permission to communicate with a targeted service, then that service may never be invoked.

\textbf{Footloose Services}: Blip envisions services which are small and accessible across the network (see Workload Element, below). Using a dynamic (continually running) and hybrid (e.g. considers multi-parametric, constraint programming based) placement algorithm, Blip will attempt to pull services toward the place of demand, minimising network latency. We refer to this as “Footloose Services”. 

\textbf{Blip Distributed Operation}: Blip application are expected to execute in a distributed way, with copies of the application’s constituent services duplicated at various points in the network. When a break in network connectivity occurs a Blip application should be able to operate in each of the resulting subnets assuming that each have copies of all the necessary constituent services. When the network heals and the two subnets are united state synchronisation will be necessary, whilst this is likely to be application-specific techniques such as those outlined in the implementation of Apache Flink \cite{c75} can be considered. In mapping this approach to Blip, each physical implementation of a Flink “stream operation” can be considered as a service executing on a Blip node. The data stream is therefore the directed data flows across the Blip graph.

\textbf{Blip Distributed Orchestration, Management and Inter-service Communication}: Current cloud management system struggle to start a virtual machine in under 3 minutes \cite{c3} (the average Azure VM takes 6 and a half minutes). Additionally they typically haven't been designed to account a wide geographic deployment of services and the latencies associated with these network distances. Therefore we need to redefine how services are deployed, and managed, and how interservice communication works. Blip envisions a distributed management framework which can span large geographic areas and still support the rapid deployment and execution of services and applications. 

\begin{figure}[thpb]
\centering
\framebox{\parbox{3in}{
\includegraphics[scale=0.57]{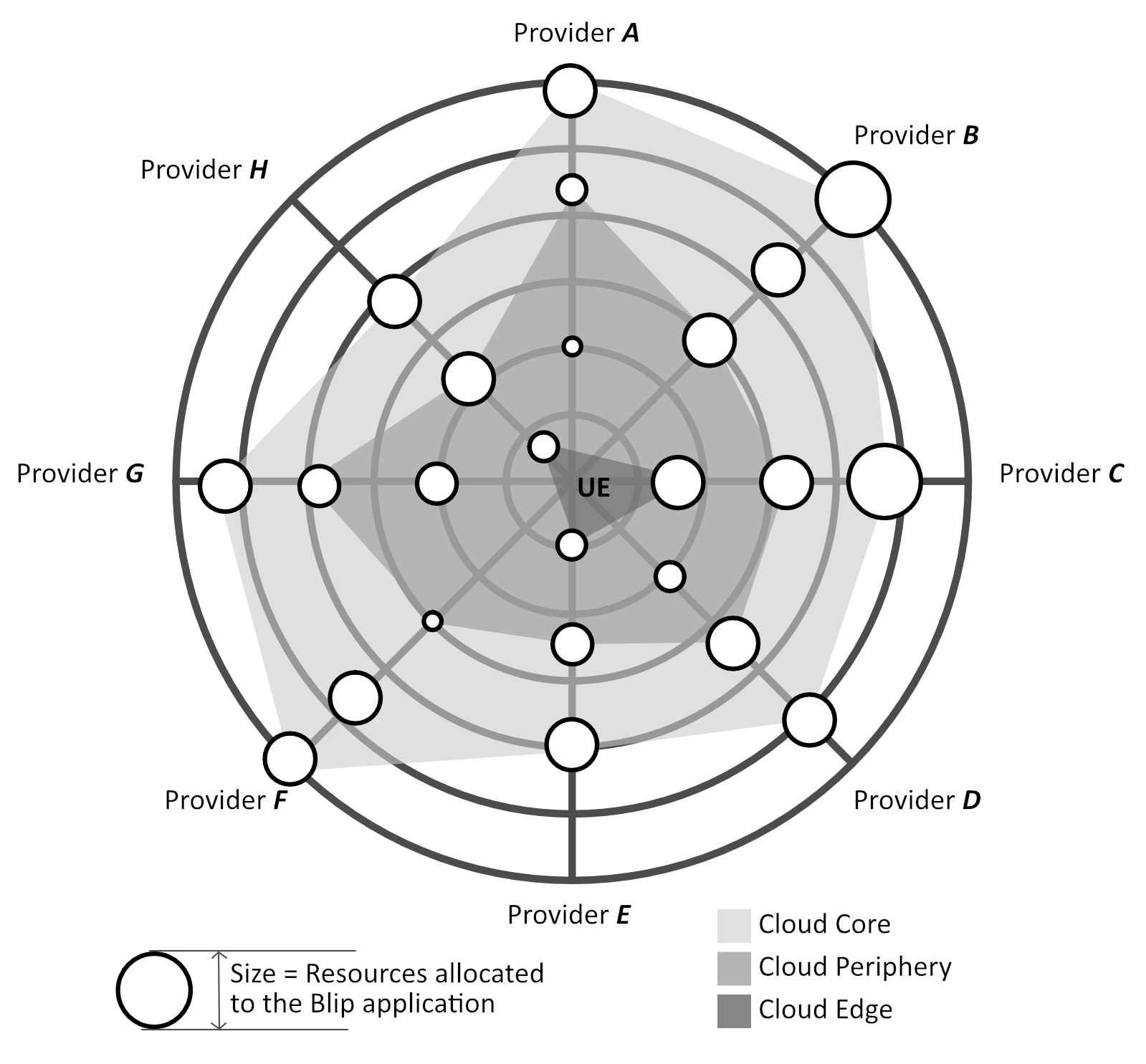}
}}

\label{radialfig}
\end{figure}

\textbf{Modeling Blip}
We can model a Blip application as a set of interconnected nodes on a graph as shown in Figure \ref{radialfig} where 1) A node is a service, part of an application and 2) an edge is a network link with particular weights, in particular captures latency.

In Figure \ref{radialfig}, the core elements of the Blip model are shown and a single but complete Blip application deployment is shown. In this depiction the Blip application is deployed across a number of geographic locations. The locations are provided by eight different providers (A-H). Each provider is mapped on its own axis. The size of the node is related to the number of resources allocated. The distance of the node from the relative origin, which in this case is a collections of User Equipment (UE; e.g. mobile phone) is the computed latency. This latency is a relative one, algorithms such as Vivaldi \cite{c2} can be used to compute this. There is also a number of cell tower planning techniques, used by telecommunications operators, that possibly could be used \cite{c63, c64, c69}. Finally, each node is grouped by category type; dark grey is the Cloud Edge, which has low latency and is near the UE, medium grey is the Cloud Periphery and light grey is the Cloud Core. We extend this graph later in section \ref{disc_conc}. We must acknowledge that the graph itself is not static. It is  constantly changing as it expands and contracts spatially and locally,  reflecting fast provisioning of resources and  the dynamic geographic movement of nodes (Footlooseness). 

Blip is intended to work across large geographic distances, with each location hosting a “Blip Stamp”.  The Blip Platform is a collection of one or more interconnected Blip "Stamps". This produces a single platform which can host workload elements in one or more geographic areas.The Blip Stamp can be anything from a single compute device, through to a large data centre. The minimum requirements (in order to run upon heterogeneous infrastructure types) for the stamp are that it is compute and storage capable, and be IP-network addressable.:

\begin{figure}[thpb]
\centering
\framebox{\parbox{3in}{

\includegraphics[scale=0.67]{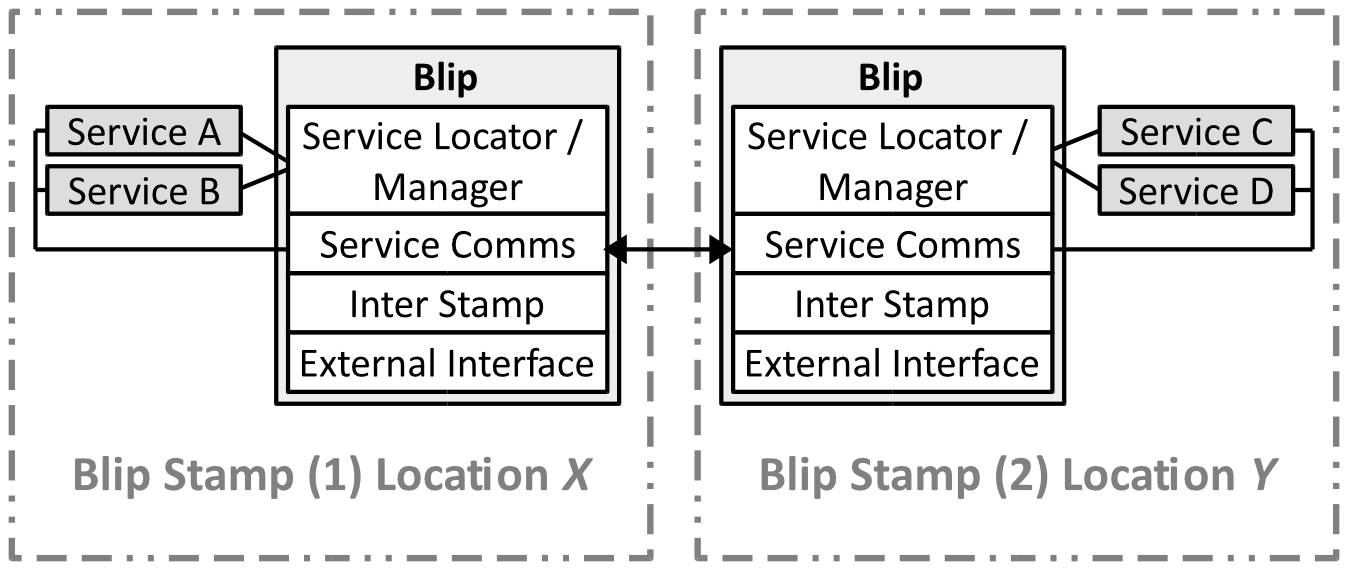}

}}

\label{figurelabel}
\end{figure}

Inside the Blip Stamp is an instance of “Blip” this provides the hosted services within each Blip Stamp the following four major functional elements. \textbf{Service Locator / Manager}. When service \textbf{\textit{A}} requests to use a second service \textbf{\textit{D}}, it is the Service Locator / Manager which handles this request. It has knowledge of where other instances of services are located and additional runtime information about their load and latency. This component can start a new instance of a service if one is not found. 
\textbf{Service Communications}. Once the desired service \textbf{\textit{D}} has been located / or started then then two services \textbf{\textit{A}} and \textbf{\textit{D}} will need to communicate with each other. This inter service communication is supported and managed by Blip.
\textbf{Inter-Stamp Communication}. Each Blip Stamp can operate on its own. However when connected to other Blip Stamps it allows blip hosted services to propagate between Stamps. Facilitating this requires that Blip be aware of what services are available to copy to the local stamp, and what services are running on which connected stamp.
\textbf{External Interface}. This provides access to Blip Exclusive Services from non-blip hosted services / applications. Communication via the External Interface is dealt with and handled in the same way as communication from an existing Blip Hosted Service.

\section{Scenarios: How Blip Helps}
\label{solution_scenario}
From the Smart Home scenario previously described, it can be seen that there are multiple hops through various networks to arrive at the target destination. The path (incl. Transmission medium) is not necessarily the major influencing factor on latency, but rather the amount of processing that happens to the packets on the path related to the request. Reducing the number of “packet processors” on the path from mobile device to light bulb is needed.

In the Blip world, the BBU would have services deployed to handle the request emanating from the mobile device. The logic would take over and request the provisioning of a pre-cached specific application gateway (JIT and WE), from the light control service provider. This gateway would be packaged as a process that is fast to instantiate and consumes minimal resources. For this WebAssemblies or Unikernels are well suited.  Once the process has started it can handle the serving of the mobile phone and directly issue the event to the bridge, which in turn turns the light off.

The Blip \textbf{Service Stamp} intercepts the call through the \textbf{Service Locator} and it finds and creates the application proxy (\textbf{WE}). The initial request to control the light still goes the conventional route to the CP, however the request details and response details are recorded. This allows future requests to be replayed. This brings new requirements of the existing system and importantly a means to store information on each stamp that is independent of the WE and persists longer than the WE’s lifetime. 

\section{Related Work}
\label{relwork}
In this section we review what currently might resemble the vision of Blip and also what technologies could be leveraged to build Blip and realise JIT and Footloose principles. There are recent architectures that only cover part of what Blip attempts to accomplish: PiCasso \cite{c77} and INDICES \cite{c10} deal with the placement of compute nearer to the end user, whereas Jitsu \cite{c27} amply demonstrates the possibility of near just-in-time service delivery. While many services have a 50ms response time, a unikernel boot within 23ms is possible. Compute Resource Technologies are those that can deliver the WE Blip principle and at speed needed for JIT services. For these we have Unikernel technologies include: MirageOS \cite{c21}, RumpKernel \cite{c22}, ClickOS \cite{c23}, Clive \cite{c25} and IncludeOS \cite{c26}. Most of these are language specific (excluding RumpKernels), however the Unik project \cite{c11} seeks to unify them. Along the lines of reducing workload footprint and language-agnostic service delivery, the WebAssembly \cite{c78} standard and MVP implementations provide a means to not only deliver an optimised workload but to almost any device that can execute the assembly. Although these WE technologies can execute the workload they too also need to be connected. This brings us to Software Defined Networking technologies. There are many SDN controllers used today, for example: NOX \cite{c28}, OpenContrail \cite{c32} and OpenDaylight \cite{c33}. However these are typical for use with VMs, IaaS deployments. Lightweight overlay network technologies (such as Weave \cite{c34} and Flannel \cite{c35}) and service meshes (such as IstIo\cite{c61}, Linkerd\cite{c62}, conduit.io\cite{c57}) are more appropriate for use within Blip. Blip will require storage over networks that connects WEs in order to persist (long-term or temporarily) data. For this centralised DBs (e.g. Postgreql, MariaDB) will not be suitable without the overhead of implementing synchronisation between DB nodes or Blip Stamps (interesting to note the approach used in \cite{c79} and DotMesh\cite{c59}). 
Of consideration is  Information Centric Networking (ICN) and Named Data Networking \cite{c36} (an approach used in PiCasso). Dynamic services are not well supported by current ICN approaches, like those delivered by Blip. CCNx \cite{c84}, an open source implementation of NDN, has limited support of general services. Having considered resource technologies, we also need to consider the orchestration of those resources. All orchestration engines are largely designed with a centralised architecture in mind (see OpenStack itself and specifically Heat \cite{c40}, Cloudify\cite{c42}, Kubernetes\cite{c43}). Unlike these, distributed orchestrators are less however included in this are \cite{c81}\cite{c80} and \cite{c82}. Netflix Conductor \cite{c83} considered the question of distributed orchestration and deemed it to be cost-prohibitive. An orchestrator needs to manage the lifecycle of the application and its resources including scaling and dynamically updating the placement of services. There is always an actuation delay to scaling. To compensate for actuation delay requires sophisticated prediction algorithms \cite{c51}. Such algorithms are usually limited to determining the number of resources, without worrying about what sizes (or types) of resources to choose \cite{c52}, which is particularly important across different Blip Stamps with different locations and resource offerings. For placement of resource assignments in networks common approaches of constraint programming \cite{c76} can be used or with stochastic placement in \cite{c53}.

\section{Discussion and Conclusion}
\label{disc_conc}
Ultimately, the concept behind both Edge and Blip is optimised and customised delivery of service but challenges exist. 

Starting with the compute capabilities required, we see a number of current day advances that still are not sufficient. Looking at the most obvious, containers, due to the extra software included in the average container the RAM footprint and resource overhead of a container is also higher than that of a unikernel. However unlike the container ecosystem developer and production support is something which is lacking from the unikernel world. If we move up we find FaaS. It has an inherent latency associated with them, which defeats the point of offering low latency services. FaaS as implemented by Amazon and OpenWhisk, only allows the functions to execute for a small time limited period, in OpenWhisk this is 1 minute by default \cite{c58} this prevents us from offering low latency long running processes at the edge. 
Looking to the networking resources, the Blip concept suggests that Blip controls inter Blip-service communication. However some services will require additional network services and may interact with non Blip services. This means that Blip will not only need to provision these processes (WEs) in remote location but also configure their networking. Blip related control and inter-blip service traffic between Blip Stamps will need to be secure and  encrypted requiring a form of accreditation or mutual trust between the two Blip Stamps. The concepts and approaches outlined by this paper address how compute can move over large geographic areas. This approach assumes that the WEs which encapsulate the moving code are stateless. However, very few applications are genuinely stateless. 
As we assume WEs to be stateless, the application or service state is stored in a clod storage service. The bottleneck for latency would move from the speed of code execution to the time taken to retrieve data from a storage location. The data would need to be pre-provisioned at the multiple geographic locations in which it is needed. It would also need to be synchronised. This is a challenge and there are some attempts to address this (e.g. NDN). Data synchronisation becomes a job of managing the small changes in data state from the last check in until the current time. This approach would not hold for all applications and further investigation is suggested in this area. 
Once the application and it's services are distributed over a wide area upon the relevant resources, and a new version of a service is published there will need to be some automated method of replacing running services and updating old cached copies. This will need to be performed while the application is still running, taking into account inter-service dependencies. Further research is needed to address how this might best be managed. Services in Blip will be created and destroyed quickly and repeatedly and in data centres and locations owned and managed by various providers. Each time they execute they will generate logging, metrics and other monitoring information which will need to be collected together and presented to the developer so that they can better understand how their application is performing. 
Importantly from a development view, debugging an application distributed across a wide geographic area, where services don't exist until they are needed is going to be a challenge. Issues with the small delays as Blip creates new instances of services can introduce random additional time delays in inter service calls. This can lead to some unexpected behaviour. 

The Radial model presented in Figure \ref{radialfig} can be further improved. There are more factors (e.g. cost, privacy) than just latency which can affect placement. The model is currently univariate oriented, it should be multivariate. The application model presented in Figure \ref{radialfig} does not include the internal dependencies between the Blip service instances. These links and the parameter between them also factor into the placement decisions and should be represented in the model. The radial model provides a snapshot of the state of the application. For this to allow continuous evaluation and movement of software across the network it needs to be computed on a continual basis. 
A distributed algorithm which provides a local, rather than globally optimised approach may offer more real-world performance and reliability. This paper excludes the definition of such an algorithm as it was realised that further research in this area is required. Additionally, the potential impact of edge on existing cloud systems has been estimated some industry people to be approximately 23\% of workload \cite{c60}, which would point to accommodating edge in application architecture but not exclusively revolving around Edge. It is for this reason why footloose services will be important, allowing for opportunistic dynamic topology changes.  Blip provides the impetus for further research in these areas and also provides an architecture upon which new technologies can be applied against. Current efforts in systems research and engineering still have to be furthered in order to address some of the needs of Edge, Blip and beyond it.

\textbf{Acknowledgements} This work is partially funded by the Swiss State Secretariat for Education, Research and Innovation (SBFI) in association with the European Union’s Horizon 2020 research and innovation programme via grant agreement \#731535, for the ElasTest project \cite{c56}. Also partially funded by this work is the mF2C project \cite{c70} under grant agreement \#730929.

\end{document}